\documentclass[twocolumn,showpacs,preprintnumbers,amsmath,amssymb,aps,prl]{revtex4}

\usepackage[pdftex]{graphicx}
\usepackage{color}
\usepackage{bm}
\usepackage{latexsym}
\usepackage{textcomp}

\def\ln{\mbox{ln}}

\renewcommand{\section}[1]{\textit{#1} --}

\usepackage{epsfig}
\usepackage{color}
\usepackage{bm}
\usepackage{latexsym}
\usepackage[utf8x]{inputenc}
\usepackage{ucs}

\begin{document}

\title{Thermal Transport in Phononic Cayley Tree Networks}
\author{H. Li, T. Kottos}
\affiliation{Department of Physics, Wesleyan University, Middletown, Connecticut 06459}
\author{B. Shapiro}
\affiliation{Technion - Israel Institute of Technology, Technion City, Haifa 32000, Israel}
\date{\today}

\begin{abstract}
We analytically investigate the heat current ${\cal I}$ and its thermal fluctuations $\Delta$ in a branching network without loops (Cayley tree). 
The network consists of two type of harmonic masses: vertex masses $M$ placed at the branching points where phononic scattering occurs 
and masses $m$ at the bonds between branching points where phonon propagation take place. The network is coupled to thermal reservoirs
consisting of one-dimensional harmonic chains of coupled masses $m$. Due to impedance missmatching phenomena, both ${\cal I}$ and 
$\Delta$, are non-monotonic functions of the mass ratio $\mu=M/m$. In particular, there are cases where they are strictly zero below some 
critical value $\mu^*$.
\end{abstract}
\maketitle

In the last two decades considerable  effort has been invested in
developing appropriately engineered structures that display novel
transport properties not found in nature. In the thermal transport
framework, this activity has recently start gaining a lot of
attention. Apart from the purely academic reasons, there are
growing practical needs emerging from the efforts of the
engineering community to manage heat transport on the nanoscale
level. Some of the targets that are within our current
nanotechnology capabilities include the generation of nanoscale
heat-voltage converters, thermal transistors and rectifiers,
nanoscale radiation detectors, and heat pumps.

Despite the considerable effort, the understanding of thermal transport possesses many challenges \cite{LLP03,D08,LRWZHL12,
COGMZ08,NGPB09,ZL10}. For example, the macroscopic laws that govern heat conduction in low dimensional systems and, in 
particular, the conditions for validity of the Fourier law remains unclear. By now it has been clarified that chaos is neither sufficient 
nor necessary for the validity of Fourier law \cite{LLP97,LCWP04,PC05}. Further research indicated the importance of the spectral 
properties of heat baths \cite{D01} and the existence of conservation laws \cite{LLP03, D08} (though see \cite{SK14}). We point 
out that although there is an established literature as far as the mean heat current is concerned, there are few results available 
about its statistics \cite{statistics1,statistics2,statistics3}.

At the same time, a variety of real structures such as biological
systems \cite{D98} and artificial networks in thin-film
transistors and nano-sensors \cite{HHG04} do not fall into the
categories of standard one- or two-dimensional lattice geometries.
Instead, they are characterized by a complex topology that,
nevertheless, can be easily realized in the laboratory
\cite{KMA05,COMZ06,PCWGD05}. It is therefore useful to employ
analytically simple models which allow us to investigate the
underlying physical mechanisms associated with heat transport in
complex networks. Along this line of thinking, 
previously, a fully connected network (each mass is  
connected to all other masses) has been studied with the help of  
random matrix theory \cite{statistics3}.

\begin{figure}
\includegraphics[width=\columnwidth]{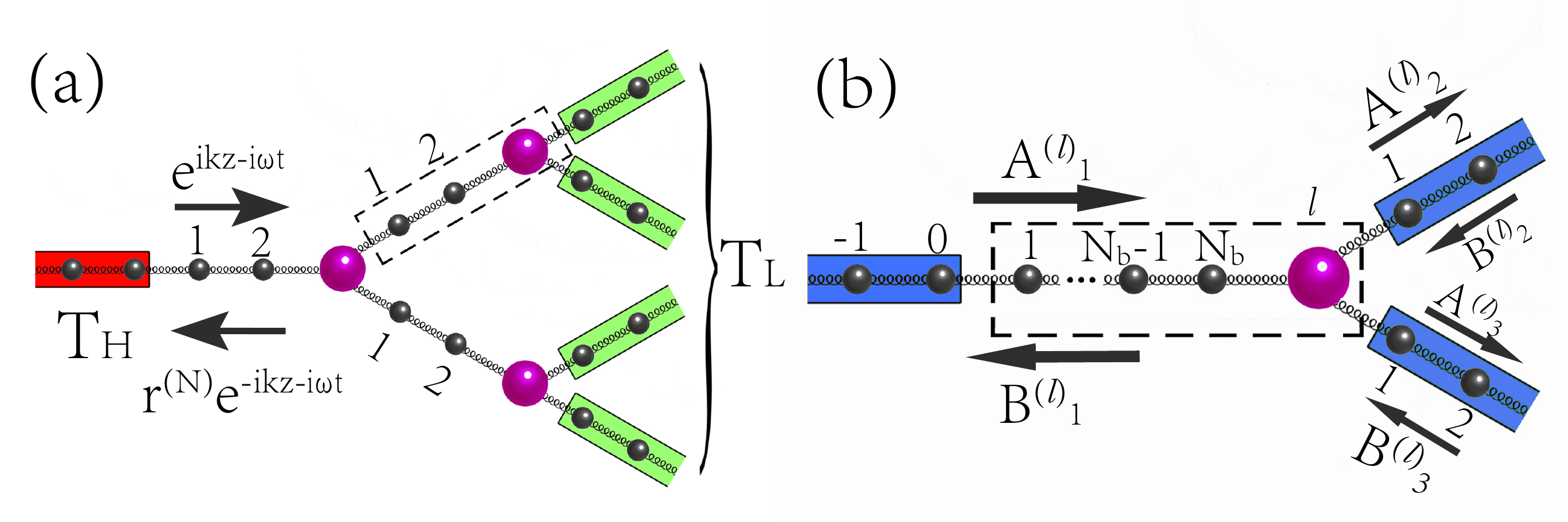}
\protect\caption{(Color online) (a) The schematic figure is a Cayley tree of generation $N=2$ with $N_{b}=2$ `small'
masses $m$ on each branch. The basic unit, is indicated with a dashed square and it consists of a branch and a vertex 
attached to the right end of the branch. A monochromatic incident wave $e^{ikz-i\omega t}$ is reflected back as 
$r^{(N)}e^{-ikz-i\omega t}$, where $r^{(N)}$ is the total reflection amplitude. For the study of thermal transport, the 
Cayley tree is connected to two heat baths at different temperatures, $T_{H}$ and $T_{L}$ respectively. The reservoirs
consist of harmonic one-dimensional atomic chains and are assumed to be in equilibrium. (b) A scattering process 
occurring at a basic unit associated with the $l$th vertex. The basic unit is connected to three semi-infinite leads (blue 
boxes). Three incoming waves from the leads are transformed to three outgoing waves at each of the leads attached to the basic unit.
\label{fig1} }
\end{figure}

In the present paper we study heat current and its thermal fluctuations in a class of harmonic mass networks which are
topologically equivalent to Cayley trees. This class of networks does not include any closed loops and has been extensively 
used in various areas of physics \cite{MP,B82}. A phononic Cayley tree consists of vertices (branching points) and branches 
in which the metric information is introduced. To be specific, each branch is an one-dimensional chain consisting of $N_{b}$ 
equal masses $m$ and $(N_{b}+1)$ springs with the same equilibrium length and spring constant $\kappa$. The vertices 
are occupied by masses $M$. The tree is characterized by its connectivity $Z$ (number of branches emanating from a vertex) 
or equivalently by the associated branching number $K=Z-1$, and its generation $N$ (number of branching repetition). For 
each generation $N$, the total number of vertices and branches are $(K^{N}-1)/(K-1)$. We will 
consider trees with $Z=3$ which are connected to two heat baths $H$ and $L$ kept at temperatures $T_{H}, T_{L}$ respectively 
with $T_H>T_L$. The bath $H$ consists of one-dimensional semi-infinite spring chain of equal masses $m$ coupled together 
with spring constants $\kappa$. The bath $L$ contains $K^{N}$ similar one-dimensional semi-infinite spring chains. An 
example of a Cayley tree for $N=2$ and $N_b=2$ is shown in Fig.\ref{fig1}(a). We find that in the large $N\rightarrow \infty$ 
limit the transmittance, as a function of frequency of the incident wave, exhibits stop bands (total reflection) and pass bands 
(only partial reflection). A direct consequence is that both the heat current ${\cal I}_{\infty}$ and its thermal fluctuations $\Delta$ 
acquire some constant value thus reflecting the ballistic nature of the heat transport in such trees. Furthermore we show that, 
due to impedance mismatch phenomena, they are non-monotonic functions of the mass ratio $\mu=M/m$: they get their 
maxima at $\mu\sim 1$ while increasing/decreasing $\mu$ leads to a decrease of their value. In particular, there are cases 
where both heat current and its fluctuations are strictly zero below some critical value $\mu^*$. Our analysis below applies 
equally well to Cayley trees with and without an on-site pinning potential $\kappa_{0}$.

{\it Theoretical Formalism -}Formally, the steady-state thermal current for the Cayley tree is obtained by 
using the Landauer-like formula \cite{D08}
\begin{align}
{\cal I}= & \int_{0}^{\infty}\frac{\mathrm{d}\omega}{2\pi}\hbar\omega {\cal T}\left(\omega\right)\left(f_{H}\left(\omega\right)-
f_{L}\left(\omega\right)\right),
\label{eq: Landauer_like}
\end{align}
with $f_{\alpha}=\left\{ \exp\left(\hbar\omega/k_{B}T_{\alpha}\right)-1\right\} ^{-1}$ being the Bose-Einstein distribution for 
the heat bath $\alpha=H,\: L$ and ${\cal T}(\omega)$ is the transmittance. The latter can be expressed via the total reflection 
amplitude $r(\omega)$ to the left reservoir as ${\cal T}(\omega)=1-|r(\omega)|^2$. The simplicity of the Cayley tree set-up 
permits us to express the full counting statistics (FCS) in terms of ${\cal T}(\omega)$. The associated steady-state cumulant 
generating function $\ln Z\left(\xi\right)\equiv\ln\left\langle e^{i\xi Q}\right\rangle$ is \cite{Li2012}
\begin{align}
\lim_{t_{M}\rightarrow\infty} & \frac{\ln Z\left(\xi\right)}{t_{M}}=-\int_{0}^{\infty}\frac{\mathrm{d}\omega}{2\pi}\ln\left\{ 1-
\mathcal{T}\left(\omega\right)F\left(\omega\right)\right\} ,
\label{eq: cumulant GF}
\end{align}
where
$F\left(\omega\right)=\left(e^{i\xi\hbar\omega}-1\right)f_{H}\left(1+f_{L}\right)+\left(e^{-i\xi\hbar\omega}-1\right)f_{L}
\left(1+f_{H}\right)$. The random variable $Q$ defines the total
amount of heat flowing out of the heat bath $H$ during the time
$t_{M}$. Note that the steady- state thermal current Eq.
\eqref{eq: Landauer_like} is related to the first cumulant of $Q$ as ${\cal I}=\lim_{t_M\rightarrow\infty}\langle Q\rangle/t_M$. Likewise,
the second cumulant gives the current noise
\begin{align}
\Delta\equiv\frac{\left\langle \left(\Delta Q\right)^{2}\right\rangle }{t_{M}}= & \int_{0}^{\infty}\frac{\mathrm{d}\omega}{2\pi}\left(\hbar\omega\right)^{2}\{ \left(f_{H}+f_{L}+2f_{H}f_{L}\right)\mathcal{T}\left(\omega\right)\nonumber\\
&+\left(f_{H}-f_{L}\right)^{2}\mathcal{T}^2\left(\omega\right)\} .
\label{fluct}
\end{align}
Thus the analysis of the FCS of heat current in a Cayley tree reduces to the study of the transmittance $\mathcal{T}
\left(\omega\right)$.

{\it Transmission Coefficient -} The analysis of the transmission coefficient of the Cayley-tree is best carried out 
using a wave-scattering approach \cite{S83,Ch90}. 

We first derive the scattering matrix for a basic scattering unit. The latter consists of a mass $M$ placed at a generic vertex $l$ 
and $N_b$ masses $m$ associated with a branch attached to the left of the vertex (see Fig. \ref{fig1}a). The equilibrium position 
of the $j$-th mass is $z_j=a j$ where $a$ is the equilibrium distance between consequent masses and $j=1,\cdots,\left(N_b+1
\right)$. Below we set $a=1$. The vertex $l$ is placed at the right end of the branch at $j=N_b+1$. The scattering problem 
associated with the basic unit is defined by attaching one semi-infinite lead with masses $m$ at the left end of the branch and two 
other identical semi-infinite leads extended at the right of the $l$-th vertex. The distance $z_j$ at the leads is always measured from 
left to right (see Fig \ref{fig1}b). The displacement of any of these masses can be expressed in terms of two counter-propagating waves
\begin{equation}
u_{n}^{(l,j)}=\left( A_n^{(l)}e^{ikz_{j}}+ B_n^{(l)}e^{-ikz_{j}}\right) e^{-i\omega t},
\label{disp}
\end{equation}
where the subindex $n=1$ indicates the branch and the associated left lead and $n=2,3$ the remaining two leads extended to 
the right of the vertex $l$. The frequencies 
$\omega$ are given by the dispersion relation $\omega=\sqrt{\frac{2\kappa}{m}\left(1-\cos k\right)+\frac{\kappa_{0}}
{m}}$, with $k\in[0,\pi]$, so that the propagating waves Eq. (\ref{disp}) satisfy the equations of motion for the masses 
$m$ at the branch and leads \cite{note1}. Furthermore Eq. (\ref{disp}) has to satisfy a consistency relation at the vertex:
\begin{equation}
\label{con}
u_{1}^{(l,N_b+1)}=u_2^{(l,0)};\quad u_{1}^{(l,N_b+1)}=u_3^{(l,0)}
\end{equation}
together with the equation of motion for the mass $M$
\begin{equation}
\label{eom}
M {\ddot u}_2^{(l,0)}=\kappa (u_2^{(l,1)}-3u_2^{(l,0)} + u_3^{(l,1)}+u_{1}^{(l,N_b)})-\kappa_0 u_2^{(l,0)}
\end{equation}
Substituting into Eqs. (\ref{con},\ref{eom}) the expressions Eq.
(\ref{disp}) we find the basic unit scattering matrix $S^{(l)}$ which
connects incoming to outgoing waves as $(B_1^{(l)}, A_2^{(l)},
A_3^{(l)})^T=S^{(l)}(A_1^{(l)}, B_2^{(l)}, B_3^{(l)})^T$:
\begin{align}
S^{(l)}= & \frac{1}{3+iF}\begin{bmatrix}-(1+iF)e^{2i\alpha} & 2e^{i\alpha} & 2e^{i\alpha}\\
2e^{i\alpha} & -\left(1+iF\right) & 2\\
2e^{i\alpha} & 2 & -\left(1+iF\right)
\end{bmatrix},\label{eq: scattering matrix}
\end{align}
where $F=\left(3-2\mu\right)\tan\frac{k}{2}+\frac{\kappa_{0}\left(1-\mu\right)}{\sin k}$, and $\alpha=k\left(1+N_{b}\right)$ 
accounts for the accumulated phase due to the wave propagation through the branch $n=1$. The scattering matrix $S^{(l)}$ 
satisfies the unitarity condition $(S^{(l)})^{\dagger}S^{(l)}=I$ and the time-reversal symmetry constrain $(S^{(l)})^{T}=S^{(l)}$.

Next, we build up a tree from many scattering units and calculate the total reflection amplitude to the left lead. To this end we connect two 
trees of $N$-generation, with identical reflection amplitudes $r^{(N)}$, into a single tree of $(N+1)$-generation. This is done with the help 
of a single vertex with scattering matrix $S^{(l)}$ Eq. (\ref{eq: scattering matrix}). The reflection amplitude $r^{(N+1)}$ of the $(N+1)$ 
generation tree can be calculated in terms of the reflection amplitudes $r^{(N)}$ and the matrix $S^{(l)}$ which connects incoming to
outgoing waves. Then, using the relations $B_{1}^{(l)}\equiv
r^{\left(\mathrm{N}+1\right)}A_{1}^{(l)},\: A_{2}^{(l)}\equiv \frac{B_{2}^{(l)}} {r^{\left(\mathrm{N}\right)}},\: A_{3}^{(l)}\equiv\frac{B_{3}^{(l)}} {r^{
\left(\mathrm{N}\right)}}$ we establish the following recursion relation
\begin{align}
r^{\left(\mathrm{N}+1\right)}= x\frac{(3-iF)r^{\left(\mathrm{N}\right)}-1-iF}{3+iF-(1-iF)r^{\left(\mathrm{N}\right)}},\,
r^{(1)}=-x\frac{1+iF}{3+iF}
\label{recursion}
\end{align}
where $x=e^{2i\alpha}$. The initial condition $r^{(1)}=S_{11}^{(l)}$ is provided by Eq.~\eqref{eq: scattering matrix}. From
Eq. (\ref{recursion}) we get
\begin{align}
r^{\left(\mathrm{N}\right)}= & \frac{2x\left(-i+F\right)}{\left(\frac{2}{-1+Y^{\mathrm{N}}}+1\right)U-(x+1)F-3i(x-1)},
\label{nGrefl}
\end{align}
where $U(\omega)=\pm\sqrt{2y}e^{i\alpha}$, $y(\omega)=\left(F^{2}-9\right)\cos2\alpha-6F\sin\left(2\alpha\right)-F^{2}+7$ and
$Y(U)=1/Y\left(-U\right)=\frac{U+F(x-1)+3i(x+1)}{-U+F(x-1)+3i(x+1)}$. The last relation can be used to show that $r^{\left(\mathrm{N}\right)}$ 
is insensitive to the choice of sign $\pm$ in $U$.

When $y<0$ the modulus of $Y$ is different from unity and Eq. (\ref{nGrefl}) converges to a  fixed point $r^*$ with $|r^*|=1$ for
$\mathrm{N}\rightarrow\infty$. In contrast, when $y>0$ we have $|Y|=1$ and Eq. (\ref{nGrefl}), does not 
converge. We conclude therefore that the convergence of the reflection amplitude $r^{(N)}$ is 
determined by the sign of the {\it band-gap parameter} $y$. 

In the frequency domain for which Eq. (\ref{nGrefl}) converges, i.e. $y<0$, we have
$\mathcal{T}^{\left(\mathrm{N \rightarrow\infty}\right)}=0$.
We refer to this frequency domain as Stop Bands (SB). The situation
is more complicated in the frequency domain for which $y>0$. In
this case $Y$ is unimodular and thus it can be written as
$Y=e^{i\varphi}$. Substituting back into Eq. (\ref{nGrefl}) we
obtain the following expression for the transmittance
\begin{equation}
\label{trans}
\mathcal{T}^{\left(\mathrm{N}\right)}=\frac{y}{y-\left(1+F^{2}\right)\left(1-\cos\left(\mathrm{N}\varphi\right)\right)}
\end{equation}
which fluctuates between a maximum value $\mathcal{T}_{max}=1$ and a minimum value $\mathcal{T}_{min}=1-\frac{1+F^{2}}
{\left(F\cos\alpha-3\sin\alpha\right)^{2}}$ which is independent of the generation number $N$. We refer to the frequency domain 
for which $y>0$ as Pass Bands (PB). Within each PB we define a smoothed version of transmittance as
\begin{equation}
\label{meanT}
\overline{\mathcal{T}}\equiv\frac{\mathcal{T}_{min}+\mathcal{T}_{max}}{2}=1-\frac{1+F^{2}}{2\left(F\cos\alpha-3\sin\alpha\right)^{2}},
\end{equation}
while within the SB we can approximate the transmittance with its asymptotic value i.e. $\overline{\cal T}=0$. We will see below that these 
approximations describe well our numerical results for the heat current and its fluctuations in the limit $N\rightarrow \infty$. The transition 
points between a PB and a SB correspond to frequencies for which $y(\omega)=0$.

\begin{figure}
\includegraphics[width=\columnwidth]{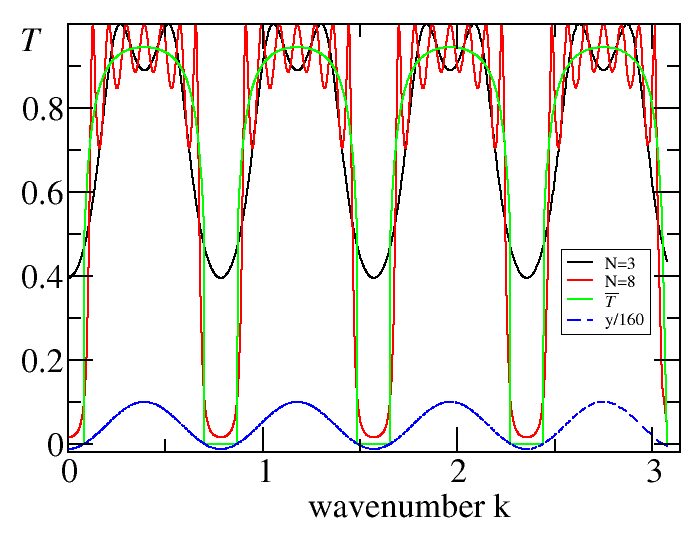}
\protect\caption{(Color online) Transmittance $\mathcal{T}^{(N)}=1-|r^{(N)}|^2$ vs wave number
$k$ for various $N$-values obtained using Eq. (\ref{nGrefl}). For comparison, we also plot the mean
value approximation $\overline{\cal T}$ and the band-gap parameter $y$ (scaled by 160 to fit the figure).
Here $m=1$, $\kappa=1$, $\mu=1.5$, $\kappa_{0}=0$ and $N_{b}=3$.
\label{Transmittance} }
\end{figure}

In Fig. \ref{Transmittance} we present two typical transmission spectra for a Cayley tree network of $N=3$ and $N=8$. We see
that they consist of alternating SB and PB as the wave number $k$ changes from $0$ to $\pi$. Armed with the knowledge about 
the transmittance
of a Cayley tree we are now ready to investigate the heat current and its fluctuations Eqs. (\ref{eq: Landauer_like},\ref{fluct}).

\begin{figure}
\includegraphics[width=\columnwidth]{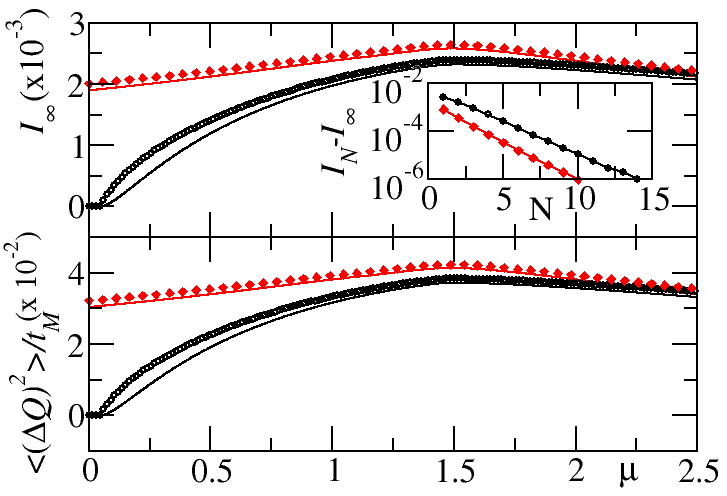}
\protect\caption{\label{fig:current_mu}(Color online) (a) The steady-state thermal current ${\cal I}\left[\hbar\frac{\kappa}{m}\right]$ 
versus the mass ratio $\mu$ for $N_b=0$ (black line), and $N_b=3$ (red line). We have used Eq. (\ref{eq: Landauer_like}) where the 
total transmittance has been calculated numerically using Eq. (\ref{nGrefl}) for $N=50$. Black circles and red diamonds indicate the 
theoretical results for $N_b=0$ and $N_b=3$ when we use the mean value approximation for the transmittance. Inset: The exponential
convergence of ${\cal I}$ with generation $N$ to its asymptotic value ${\cal I}_{\infty}$.Two typical cases with $\mu=0.05$ for the 
$N_b=0$ (black line) and $N_b=3$ (red line). (b) The same as in (a) but now for the thermal fluctuations $\Delta$. The parameters used 
are $T_{H}=0.41\,\frac{\hbar}{k_{B}} \sqrt{\frac{\kappa}{m}}$, $T_{R}=0.39\,\frac{\hbar}{k_{B}}\sqrt{\frac{\kappa}{m}},$ $\kappa_{0}=0$.
}
\label{fig3}
\end{figure}

{\it Heat Current and its fluctuations -} First we consider the scaling of the steady-state thermal current ${\cal I}_N$, calculated
using Eq. (\ref{eq: Landauer_like}), with respect to the tree generation $N$. A detailed scaling analysis indicates that ${\cal I}_{N}$ 
converges to its asymptotic value exponentially fast i.e. ${\cal I}_{N}-{\cal I}_{N\rightarrow\infty}\sim\exp(-\gamma N)$ (see inset 
of Fig. \ref{fig:current_mu}a). Below we will be using $N=50$ as a good approximation for the asymptotic heat current ${\cal I}_{\infty}$.

In the main panels of Fig. \ref{fig:current_mu} we report the
dependence of ${\cal I}_{\infty}$ and its fluctuations $\Delta$,
Eqs. (\ref{eq: Landauer_like},\ref{fluct}), on the mass ratio
$\mu=M/m$ for two representative values of the number of masses
$N_b=0,3$ at the branches. The solid lines in these figures
correspond to the numerical results obtained from Eq. (\ref{nGrefl}). For comparison we 
plot the theoretical predictions (symbols)  associated with the
approximated expression for the transmittance $\overline{\cal T}(\omega)$
given in Eq. (\ref{meanT}). 

Generally, the asymptotic value ${\cal I}_{\infty}$ is nonzero, reflecting the ballistic nature of the thermal transport across
a Cayley tree. We find that ${\cal I}_{\infty}$ is a non-monotonic function of the mass ratio $\mu$. Indeed, in the two limiting 
cases of $\mu\ll 1$ and $\mu\gg 1$ there is a considerable impedance miss-matching between the masses $m$ of the 
leads attached to the reservoirs and the mass $M$ at the vertices. This impedance miss-match is, in turn, responsible for the 
reflection of the energy flowing from the lead to the tree and thus for the decrease of the heat current. As the mass ratio 
$\mu$ approaches unity the impedance matching is restored and heat current flows from the hot reservoir towards the cold 
reservoirs via the Cayley tree.

Although the $N_b=3$ case is representative of the dependence of heat current on $\mu$, the $N_b=0$ case shows some
non-generic features. Specifically, we find that ${\cal I}_{\infty}$ vanishes for a restricted parameter range $0\leq \mu\leq
\mu^*$. The critical value of the mass ratio can be evaluated exactly once we take into consideration the band-gap structure
of the transmittance spectrum ${\cal T}(\omega)$. Specifically for $N_b=0$, we find that $y(\omega)<0$, for any $\omega$
given by the dispersion relation as long as
\begin{equation}
\label{mu}
\mu\leq \mu^*= \frac{\left(3-2\sqrt{2}\right)\kappa+\kappa_{0}}{4\kappa+\kappa_{0}}.
\end{equation}
In this case we only have a SB with ${\cal T}(\omega)=0$ and thus ${\cal I}_{\infty}=0$.  For
Cayley trees with $N_{b}\neq0$, the band-gap parameter $y(\omega)$ changes sign at least once as the wave number varies
in the interval $[0,\pi]$. Therefore the asymptotic current is different from zero for any finite $\mu$.

\begin{figure}
\includegraphics[width=\columnwidth]{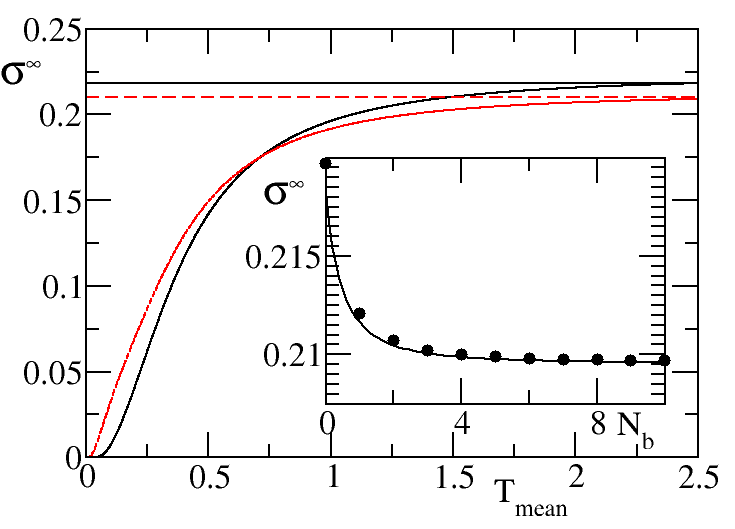}
\protect\caption{(Color online) Numerical values of conductance $\sigma^{\infty} \left[k_{B}\sqrt{\frac{\kappa}{m}}\right]$
versus temperature $T_{\rm mean}=(T_H+T_L)/2$ $\left[\frac{\hbar}{k_{B}}\sqrt{\frac{\kappa}{m}}\right]$
for $N_{b}=0$ (black solid line) and $N_b=3$ (red dashed line). The horizontal lines (of the same
type and color) are the classical results of Eq. (\ref{clas}). Inset: Numerical values (black circles) and
classical values Eq. (\ref{clas}) (solid black line) of conductance versus $N_b$. The numerical results
corresponds to $N=50$ and we have used Eq. (\ref{eq: Landauer_like}) where the total transmittance has
been calculated using Eq. (\ref{nGrefl}). Other parameters are $\mu=1.5$, $\kappa_{0}=0$. For the inset
we have used $T_{\rm mean}=3 \left[\frac{\hbar}{k_{B}}\sqrt{\frac{\kappa}{m}}\right]$.
\label{fig:conductance}}
\end{figure}

Finally, as an example, we report in Fig.~\ref{fig:conductance} the temperature-dependence of the asymptotic conductance
$\sigma^{(N\rightarrow\infty)}\equiv \lim_{T_H\rightarrow T_L} {\cal I}(T_H,T_L)/(T_H-T_L)$ for $\kappa_{0}=0$ and $\mu
=\frac{3}{2}$. In the high-temperature limit, the asymptotic current approaches its classical value leading to a saturation of 
the conductance. It is thus instructive to calculate the classical limit of ${\cal I}_{\infty}$. In this case the Landauer-like formula 
Eq.~\eqref{eq: Landauer_like} reduces to
\begin{align}
{\cal I}_{\infty}^{Cl}= & \int_{0}^{\infty}\frac{\mathrm{d}\omega}{2\pi}\mathcal{T}\left[\omega\right]k_{B}\left(T_{H}-T_{L}\right)
\label{CC}
\end{align}
which can be analytically evaluated using the mean value approximation for the transmittance Eq. (\ref{meanT}) within the PB 
frequency range. The PBs frequency range can be obtained from the analysis of the band-gap parameter $y$. The corresponding 
wave-numbers take values in the interval $k\in\left[\frac{\arccos\frac{7}{9}+2\pi n}{2\left(N_b+1\right)},\:\frac{2\pi\left(n+1\right)
-\arccos\frac{7}{9}}{2\left(N_b+1\right)}\right]$ for $n=0,\,1,\ldots,N_b$. The resulting expression for the classical current is then
\begin{align}
I_{\infty}^{Cl}= & \frac{k_{B}\sqrt{\kappa}}{\pi\sqrt{m}}\frac{\sin\left(\frac{\delta}{4\left(N_b+1\right)}\right)}{\sin\left(\frac{\pi}{4\left(N_b+1\right)}\right)}
\mathcal{\overline{T}}\Delta T\xrightarrow{N_b\rightarrow \infty}\frac{k_{B}\sqrt{\kappa}\delta}{\sqrt{m}\pi^{2}}
\mathcal{\overline{T}}\Delta T
\label{clas}
\end{align}
where $\delta=\pi-\arccos\frac{7}{9}$ and $\Delta T= T_H-T_L$. In Eq. (\ref{clas}) we have used the additional simplification
that $\overline{\cal T}$ is nearly a constant in all PBs.

\section{Conclusion}
We study heat transport through a Cayley tree. The tree is built out of masses $M$ (vertices) connected by branches which consist of masses $m$  
linked by identical springs. First we calculated transmission of phonon waves through the tree and show that, depending on the frequency,  
waves are either totally reflected (stop band) or get partially transmitted (pass band). Then we studied the heat current through the structure 
and show that, in the limit of an infinite tree, both the average current and its variance approach a well defined limit which, in some particular 
cases, can be strictly zero.


\begin{thebibliography}{99}
\bibitem{note1}We note that for the special case $N_b=0$, the branches does not contain any small mass `$m$' and the prescribed
dispersion relation is simply an ansatz for a unified treatment, which will be justified by introducing the branch-dependent amplitudes
$A_n,B_n$ to satisfy all the required equations of motion.
\end{thebibliography}
\end{document}